# Long-Term therapeutic effects of Katona therapy in moderate-to-severe perinatal brain damage.

Manuel Hinojosa-Rodríguez[a], José Oliver De Leo-Jiménez[a], María Elena Juárez-Colín[a], Eduardo González-Moreira[a], Carlos Sair Flores-Bautista[a], Thalía Harmony[a *]

[a] Unidad de Investigación en Neurodesarrollo, Departamento de Neurobiología Conductual y Cognitiva, Instituto de Neurobiología, UNAM, Campus Juriquilla, México.

First author: Manuel Hinojosa-Rodríguez at Unidad de Investigación en Neurodesarrollo, Departamento de Neurobiología Conductual y Cognitiva, Instituto de Neurobiología, Universidad Nacional Autónoma de México, Campus Juriquilla, México, Phone: +52 442 3374333.
E-mail: manuelhinojosa@comunidad.unam.mx; manuel_hinojosa_rodriguez@hotmail.com

*****Corresponding author: Thalía Harmony at Unidad de Investigación en Neurodesarrollo, Departamento de Neurobiología Conductual y Cognitiva, Instituto de Neurobiología, Universidad Nacional Autónoma de México, Campus Juriquilla, México, Phone: +52 442 1926101 ext 112.
E-mail: thaliah@unam.mx

## Highlights

- First multimodal study of Katona therapy effects in PBD with long-term follow-up.
- MRI and Katona evaluations allow early detection of cerebral palsy.
- Katona therapy decreases the severity of motor disability in moderate-to-severe PBD.
- Katona methodology for early detection can be applied in developing countries.

**Abstract**

*Aim:* To determine the long-term efficacy of Katona therapy and early rehabilitation of infants with moderate-to-severe perinatal brain damage (PBD).

*Methods:* Thirty-two participants were recruited (7-to-16 years) and divided into 3 groups: one Healthy group (n = 11), one group with PBD treated with Katona methodology from 2 months of corrected age, and with long-term follow-up (n = 12), and one group with PBD but without treatment in the first year of life due to late diagnosis of PBD (n = 9). Neuropediatric evaluations, motor evoked potentials (MEPs) and magnetic resonance images (MRI) were made. The PBD groups were matched by severity and topography of lesion.

*Results:* The patients treated with Katona had better motor performance when compared to patients without early treatment (Gross Motor Function Classification System levels; 75% of Katona group were classified in levels I and II and 78% of patients without early treatment were classified in levels III and IV). Furthermore, independent *k*-means cluster analyses of MRI, MEPs, and neuropediatric evaluations data were performed. Katona and non-treated early groups were classified in the same MRI cluster which is the expected for patient's population with PBD. However, in MEPs and neuropediatric evaluations clustering, the 67% of Katona group were assigning into Healthy group showing the impact of Katona therapy over the patients treated with it. These results highlight the Katona therapy benefits in early rehabilitation of infants with moderate-to-severe PBD.

*Conclusions:* Katona therapy and early rehabilitation have an important therapeutic effect in infants with moderate-to-severe PBD by decreasing the severity of motor disability in later stages of life.

**Keywords:** perinatal brain damage; Katona therapy; early treatment; cerebral palsy; magnetic resonance imaging; motor evoked potentials.

## Introduction

Perinatal brain damage (PBD) is a heterogeneous neuropathological spectrum that affects the encephalic white and/or gray matter, which can occur between the 20th week of fetal life and the 28th postnatal day [1,2]. The etiology of PBD is multifactorial and different physiopathological mechanisms can be involved, such as hypoxia, ischemia, inflammation, and/or infections [3]. The incidence of neonatal encephalopathy is 8.5 per 1000 live births, in which 50 to 80% of cases have a hypoxic-ischemic history [4]. The diagnosis and severity of PBD can be determined by indirect approaches, for instance, clinical scoring systems, biomarkers, and biochemical assessment [5]. However, magnetic resonance imaging (MRI) is the most sensitive technique that allows delineating accurately the topography and severity of the brain lesion in the neonate and infant. In this sense, the topography and severity of PBD are decisive in the establishment of short and long-term sequelae, regardless of the neuropathological pattern [6]. PBD has a high variety of adverse outcomes depending on the topography and severity of brain lesion, ranging from mild abnormalities in infancy (e.g., delay in acquiring developmental milestones) to severe neurological sequelae that cause permanent disabilities. Cerebral palsy is the main cause of childhood motor disability and may be the most serious long-term sequela of moderate-to-severe PBD [7]. However, as brain plasticity is greater in the developing than in the adult brain, timely diagnosis and early intervention are crucial.

Katona neurohabilitation therapy is a therapeutic treatment that promotes the psychomotor and cognitive capabilities that neonates and infants have not yet developed [8,9]. Katona methodology is both diagnostic and therapeutic. Katona assessment and treatment are performed with maneuvers taking into account muscle tone, symmetry of the hemibodies, attention, auditory monitoring, eye tracking, and neurological signs of alarm. Katona assessment and therapy is also based on early integrated complex movements, triggered by different head positions, the “elementary neuromotor patterns”. These patterns are a series of complex and continuous movements (for a detailed review about Katona methodology see [10]).

The main objective of this study is to establish the long-term efficacy of Katona therapy and early rehabilitation of infants with moderate-to-severe PBD. Our hypothesis, in this study, is that infants with early diagnosis and treatment with Katona methodology, will have less severe neurological sequelae observed years after in childhood and adolescence, than infants with similar brain lesions but that did not receive early treatments.

## Material and methods

The Ethics Committee of the Instituto de Neurobiología of the Universidad Nacional Autónoma de México (UNAM) approved this study, which also complies with the Ethical Principles for Medical Research Involving Human Subjects established by the Declaration of Helsinki. Informed written parental consent for participation in this study was obtained for all participants.

### *Participants*

Thirty-two children and adolescents were recruited (14 females; range 7-to-16 years) for this study. Twenty-three of them were recruited from a total of 1,694 participants evaluated in the Unit for Neurodevelopmental Research at the UNAM during the period 2001 to 2018. Three groups of children and adolescents were considered according to meet the following inclusion criteria: clinical history (e.g., multiple risk factors for PBD) and structural MRI compatible with moderate-to-severe PBD (or normal for Healthy group) when they started the treatment. i) Healthy group, with eleven right-handed participants (4 females) between 8 and 11 years of age, all of them born at term (range 37-to-41 gestational weeks), without neurologic or neuromotor illness, without risk factors for PBD, and with normal structural MRI and age-appropriate motor performance.

ii) Katona group with twelve participants (6 females) between 7 and 16 years of age, nine patients with premature birth (range 27-to-36 gestational weeks), with multiple risk factors for PBD, MRI compatible with moderate-to-severe PBD, who have received Katona treatment and early rehabilitation since 2 months of corrected age (or before), and with a 7-to-12 year follow-up in the Unit for Neurodevelopmental Research.

iii) Group treated late (without Katona therapy) with nine patients (4 females) between 7 and 16 years of age, all of them born prematurely (range 27-to-33 gestational weeks), with multiple risk factors for PBD, MRI compatible with moderate-to-severe PBD, but without formal treatment in the first 12-to-18 months of life due to late diagnosis. All patients of this group were recruited in public health centers in Mexico with late diagnosis of PBD (≥12 months of corrected age) or spastic cerebral palsy diagnosis after 18-to-24 months of age.

The PBD diagnosis was made in accordance with neuropediatric and MRI standards [5,6,11,12]. Group treated late was considered as a Control group since Katona therapy and early rehabilitation have proved to be useful [7,10,13,14]. Therefore, according to the declaration of Helsinki, no control clinical trials without treatment are permitted [15]. Participants with clinical history of traumatic brain injury, brain malformations, cardiovascular illness, and/or genetic pathology associated with brain abnormalities were excluded from this study.

***Evaluations***

Neuropediatric evaluations, specific motor performance tests, structural and quantitative MRI and motor evoked potentials (MEPs) by transcranial magnetic stimulation were made: *a) Neuropediatric evaluation and motor performance tests:* The clinical pediatric and neuropediatric examinations were made by physicians with wide experience in the field. The specific motor performance tests are: range of motion (ROM), muscle strength scale (Daniels scale), muscle tone (Ashworth scale), deep tendon reflex (DTR scale), manual ability (MACS; Manual Ability Classification System), and gross motor performance (GMFCS; Gross Motor Function Classification System).

*b*) *Structural magnetic resonance imaging (structural MRI):* Brain MRI studies in infancy stage with early PBD detection aim were only obtained for the Katona group (only group with early diagnosis and longitudinal follow-up) and were acquired with a 1.0 Tesla MRI scanner (Philips Medical Systems, Best, Netherlands). Infant brains were scanned during natural sleep and participants used ear protection. Structural images included sagittal T1-weighted and axial T2-weighted. Brain MRI studies for all participants in a range 7-to-16 years (transversal analysis) were acquired with a 3.0 Tesla MRI scanner (General Electric Healthcare, Milwaukee, Wisconsin, US). All participants were awake and used ear protection. Structural images included: coronal 3D T1-weighted SPGR, coronal 3D T2-weighted SE, axial T2-weighted FSE, 3D-TOF SPGR and diffusion tensor imaging. Time acquisition for these five sequences was approximately 22 minutes. MRI parameters can be found in the supplementary material. Type of brain lesion: brain MRI studies were categorized according to neuroradiological patterns and perinatal clinical history, which correlates with the etiology, physiopathological mechanisms, and temporality of the lesion [6,11]. All neuropathological patterns detected in this sample indicate that the timing of the brain lesion occurred in the perinatal stage, according to Krägeloh-Mann and Horber classification [12]. Grade of severity of PBD: brain MRI studies in infancy stage were classified according to white and gray matter abnormality scales [16]. All patients showed moderate-to-severe white/gray matter abnormalities. Brain MRI studies of children and adolescents were classified according to severity for white matter injury in cerebral palsy [17]. The PBD groups were matched by severity and topography of lesion, namely, moderate-to-severe brain damage and injury in the territory of the corticospinal tract (motor cortex, superior corona radiata, posterior limb of internal capsule [PLIC], midbrain peduncle, basilar region of pons, and/or pyramid of medulla oblongata) occurred in perinatal stage.

*c) Quantitative magnetic resonance imaging (quantitative MRI):* The volumes of the lateral ventricles and corpus callosum were obtained, as well as the areas of the PLIC. Manual segmentation of the corpus callosum and lateral ventricles was performed from coronal 3D T1-weighted SPGR sequence. Subsequently, estimates of the volumes and 3D reconstruction were made to verify the anatomy. The areas of the PLIC were obtained from diffusion RGB-color maps. Detailed segmentation methodology can be found in the supplementary material.

*d) Motor evoked potentials (MEPs):* The MEPs are the result of averaging the compound muscle action potentials registered from the target muscle during transcranial magnetic stimulation of the corresponding primary motor (M1) cortex. The safety, ethical considerations, and application guidelines for the use of transcranial magnetic stimulation in clinical practice and research were carried out according to Rossi et al., 2009 [18]. Two surface AgCl electrodes were positioned over tibialis anterior in a belly-tendon classic montage and connected to an amplifier. Main characteristics of the amplifier are a gain of 1000 dB; low-cut filters, 10 Hz; high filters, 3000 Hz. Electromyography was recorded during MEPs acquisition (Neuronic Mexicana, S.A.; Mexico). Single-pulses of 5 Hz were delivered with double cone coil (Super-Rapid, Magstim Company, Whitland, United Kingdom) over M1 cortex of the tibialis anterior. The M1 cortex detection was realized using single pulses at 80 to 100% of maximum stimulator output and the subsequent record of compound motor action potential in contralateral target muscle at rest. The "hotspot" area was defined as the motor cortical area that triggers a compound motor action potential with greater amplitude, most consistent MEP by cortical mapping. Motor threshold was defined as minimum stimulus that produces a small compound motor action potential, equal to or slightly greater than 50 μV of amplitude in the resting muscles in at least half of 10 consecutive trials [19]. MEPs of contralateral muscle to cortical stimulation were acquired at 120% of motor threshold or 100% of maximum stimulator output (in cases with motor threshold ≥85% of maximum stimulator output). Failure to obtain MEPs was considered as no response to 20-to-30 stimuli at maximum stimulator output [20].

*e) Motor evoked potentials (MEPs) parameters:* We acquired the motor conduction time parameters. Total and peripheral motor conduction times were defined as the time between the artifact of stimulation at the onset of compound muscle action potential and cortical and spinal stimulation, respectively. To obtain peripheral motor conduction time, spinal stimulation was performed by placing the focal coil at the level of the 5th lumbar vertebra. We calculate the central motor conduction time by subtracting peripheral motor conduction time from total motor conduction time. The conduction velocity was defined as the distance between two points (e.g., distance between vertex to tibialis anterior) divided by motor conduction time or latency (expressed in m/s). To calculate the conduction velocity parameters, were measured the distance from vertex to tibialis anterior (total conduction velocity), distance from vertex to spinal stimulation place (central conduction velocity) and distance from spinal stimulation place to tibialis anterior (peripheral conduction velocity).

### *Katona methodology*

Katona methodology and preliminary results have been described in previous papers (see Harmony et al., 2016 [10], Harmony 2017 [14] and Garófalo-Gómez et al., 2019 [13]). However, this is the first study with a multimodal evaluation of long-term therapeutic effects of Katona therapy and early rehabilitation in children and adolescents with moderate-to-severe PBD.

### *Statistical analysis*

In this paper we use the well-known *k*-means algorithm since it has proved to be a reliable and simple tool for data mining [21]. Absence of normality and heteroscedasticity were demonstrated by Anderson-Darling test and Cumulative Distribution Function of Chi Square, respectively. Comparisons between groups (gestational age, weight at birth, Apgar score at five minutes, and age during assessments) were performed using Kruskal-Wallis test. A $p$-value $<0.05$ was considered statistically significant.

## Results

The results of the comparisons of demographic variables between groups were as follows: gestational age, weight at birth, and 5-minute Apgar score were significantly different between Healthy group and Katona group. Gestational age and weight at birth were significantly different between Healthy group and Control group ($p = 0.06$ for 5-minute Apgar score comparison). No significant differences were found between Katona and Control groups when comparing gestational age ($p = 0.09$), weight at birth ($p = 0.11$), 5-minute Apgar score ($p = 0.67$), and age during assessments ($p = 0.75$). Main demographic characteristics of groups are in Table 1.

### *Neuropediatric evaluation and motor performance tests*

Table 2 shows the results of neuropediatric evaluations and specific motor performance tests of each participant. The patients treated early with Katona had better motor performance when compared to patients without early treatment. We found that 9 out of the 12 (75%) children and adolescents treated with Katona could move without the help of assistive devices; these 9 children had levels I and II according to GMFCS, 2 children (17%) in level III, and 1 (8%) in level V. In contrast, 7 out of the 9 (78%) patients with PBD who did not receive early treatment were classified in levels III and IV of GMFCS and only 2 (22%) in levels I and II. Interestingly, two patients (17%) treated early with Katona therapy had a normal motor performance; conversely, neither patient in the late treatment group had normal outcomes. The motor sequelae in the Katona group (12 patients = 100%) were hemiparesis in 6 (50%), triparesis in 2 (17%), monoparesis in 1 (8%), and quadriparesis in 1 (8%). Instead, the most severe motor sequelae were more frequent in the late treated group (9 patients = 100%); quadriparesis in 4 (45%), triparesis in 2 (22%), hemiparesis in 2 (22%), and paraparesis in 1 (11%).

***Magnetic resonance imaging (MRI)***

In relation to the MRI findings, Table 3 shows detailed characteristics of cerebral white/gray matter lesions, cerebellum abnormalities, grade of severity of injury, and neuropathological pattern. The most frequent pathologies in Katona group and Control group were: periventricular leukomalacia (PVL), arterial ischemic stroke, periventricular hemorrhagic infarction, and mixed patterns, as the PVL in concomitance with periventricular hemorrhagic infarction (Figure 1). Quantitative MRI also evidences structural alterations in both PBD groups with diminished volumes of the corpus callosum, increased volume of the lateral ventricles (Figure 2), as well as reduction of the PLIC area. The volumetric estimations of corpus callosum and lateral ventricles of all participants were made ($n$ = 32). The areas measured of the PLIC of 31 participants were obtained, although it was not possible to obtain a diffusion RGB color map of the one patient in the Katona group. Quantitative MRI results can be found in Table 4.

***Motor evoked potentials (MEPs)***

The total conduction velocity and conduction velocity of the corticospinal tract were higher in the patients treated early with Katona methodology in relation to children and adolescent without Katona therapy (Control group). Furthermore, the MEPs parameters of the Katona group show similar values to Healthy group. In this sense, the mean of the total conduction velocity from the right hemisphere to left tibialis anterior of Katona group was 59 m/s *vs* 52.8 m/s of the patients with late treatment (reference of Healthy group 63.9 m/s). The mean of the total conduction velocity from the left hemisphere to right tibialis anterior of Katona group was 57.9 m/s *vs* 49.7 m/s (Healthy group 64 m/s). Corticospinal tract velocity showed a similar trend (Healthy group > Katona group > Control group); right corticospinal velocity 52.2/46.7/39.9 m/s, and left corticospinal velocity 51.8/45.4/34.9 m/s. The mean motor threshold for right M1 cortex in Katona group was 74% of maximum stimulator output *vs* 76% of the Control group (56% motor threshold of Healthy group) and mean motor threshold for left M1 cortex in Katona group was 74% *vs* 75% (59% on Healthy group). The MEPs of 30 participants were obtained, due to failure in obtaining MEPs for two patients in the group treated lately (no response to 20-to-30 stimuli).

### *K-means cluster analysis*

The *k*-means algorithm for clustering data was applied to three different data bases: i) quantitative MRI (topography of lesion, corpus callosum volume, lateral ventricles volumes, and PLIC areas), ii) MEPs (motor threshold and conduction velocity parameters), and iii) neuropediatric evaluation/motor performance tests (ROM, Daniels scale, Ashworth scale, DTR scale and MACS). For classification purposes, the thirty-two children and adolescents recruited in this study were sorted into three classes defined by healthy participants (C1), patients that received Katona therapy (C2), and patients that no received Katona therapy (C3). Table 5 shows the confusion matrix for clustering results based on *k*-means algorithm.

## Discussion

This is the first multimodal study that provides evidence of long-term therapeutic effects of Katona therapy and early rehabilitation in infants with moderate-to-severe PDB. Katona and Control groups were matched by severity and topography of lesion, furthermore, no significant differences between these groups were found when comparing gestational age, weight at birth, and Apgar score at five minutes. Interestingly, both groups were classified in the same MRI cluster by *k*-means analysis, indicating statistically, and not only by MRI assessment, that patients of both groups have similar attributes. However, 66.6% of Katona group were misclassified as Healthy group for MEPs cluster and motor performance; both clusters with similar parameters between the groups showing the influence of Katona therapy over the patient group. Also note that for Control group the misclassification for MEPs and motor performance evaluation mainly remain on the patient section (Katona group and Control Group). This is not the first report that early intervention seems to improve motor outcomes [10, 22, 23], however, it is the first report that such favorable results are associated with Katona therapy 10 years later.

The improvement in motor performance of the Katona group may be explained by plastic changes of the undamaged or less affected motor cortex and/or corticospinal tract, with possible compensatory modifications in their development/ontogeny [24]. The critical developmental period of the motor system is around the first 12-to-18 months of life, mainly with degeneration or establishment of fast conducting ipsilateral fibers of corticospinal tracts [25]. We hypothesized that intense repetition of Katona maneuvers in very early stages of life produces a reorganization of the motor system in the brain. Another factor, as McCoy and his coworker reported, Katona therapy is also family centered [26,27]. This work has some limitations; maybe the most relevant is the absence of a control group with pre/post intervention studies, which for ethical considerations was discarded in the study design [15]. Other limitations are the sample size and the variability of treatments and/or amount of therapy applied to the group with late diagnosis of PBD. However, the results reached on this study, both subjective and objective, lays the foundation for a future study in a large sample evaluating the benefits of Katona therapy in the long-term evolution of children with PBD.

## Conclusions

Early detection in infants younger than 2 months of corrected age with PBD or perinatal risk factors that are highly probable to result in a motor disability or cerebral palsy is possible with multidisciplinary evaluations, such as Katona methodology and MRI, even in developing countries such as Mexico. This system for very early PBD detection is viable and can be applied in many other countries. On the other hand, early treatment of these infants with Katona neurohabilitation procedure may prevent serious motor disabilities. Katona therapy and early rehabilitation have an important long-term therapeutic effect in moderate-to-severe PBD by decreasing the severity of motor disability in childhood and adolescence.

## Declarations of interest

None.

## Funding source

This work was partially supported by grants of CONACYT No. 4971 and PAPIIT IN200917 project from DGAPA, UNAM.

## ACKNOWLEDGEMENTS

Manuel Hinojosa Rodríguez is a doctoral student from Programa de Doctorado en Ciencias Biomédicas, Universidad Nacional Autónoma de México (UNAM) and received fellowship 304834 from CONACYT. The authors acknowledge Ms. Teresa Álvarez-Vázquez, Ph.D. Juan José Ortiz-Retana and B.BSc Isabel Suárez-López for their technical support. Authors also acknowledge Ph.D. Michael Jeziorski for the revision of the English version of the manuscript.

**TABLE 1. Main Demographic Characteristics.**

| Group | Number of Participants | Gender | | Gestational Age (weeks) | | Weight at Birth (grams) | | Age During Assessments (years) | |
|---|---|---|---|---|---|---|---|---|---|
| | | F | M | Mean | SD | Mean | SD | Mean | SD |
| Healthy | 11 | 4 | 7 | 39.1 | 1.2 | 3,223.6 | 319.6 | 9.8 | 1.2 |
| Katona (early treatment) | 12 | 6 | 6 | 34.1 | 3.7 | 2,203.3 | 880.8 | 10.7 | 3.1 |
| Control (late treatment) | 9 | 4 | 5 | 30.8 | 2.3 | 1,485.5 | 398.3 | 11.1 | 2.9 |

**TABLE 2. Neuropediatric Evaluations and Specific Motor Performance Tests**

| Group | | Range of Motion (ROM) | Daniels Scale | Ashworth Scale | Deep Tendon Reflex | Manual Ability Classification System (MACS) | GMFCS Palisano Scale | Diagnosis |
|---|---|---|---|---|---|---|---|---|
| | | Extremity: Upper Right / Upper Left / Lower Right / Lower Left | | | | Right / Left | | |
| Healthy | H1 | C / C / C / C | 4 / 4 / 4 / 4 | 0 / 0 / 0 / 0 | ++ / ++ / ++ / ++ | 1 / 1 | N/A | Normal |
| | H2 | C / C / C / C | 4 / 4 / 4 / 4 | 0 / 0 / 0 / 0 | ++ / ++ / ++ / ++ | 1 / 1 | N/A | Normal |
| | H3 | C / C / C / C | 4+ / 4+ / 4+ / 4+ | 0 / 0 / 0 / 0 | ++ / ++ / ++ / ++ | 1 / 1 | N/A | Normal |
| | H4 | C / C / C / C | 4+ / 4+ / 4+ / 4+ | 0 / 0 / 0 / 0 | ++ / ++ / ++ / ++ | 1 / 1 | N/A | Normal |
| | H5 | C / C / C / C | 4 / 4 / 4 / 4 | 0 / 0 / 0 / 0 | ++ / ++ / ++ / ++ | 1 / 1 | N/A | Normal |
| | H6 | C / C / C / C | 4 / 4 / 4 / 4 | 0 / 0 / 0 / 0 | ++ / ++ / ++ / ++ | 1 / 1 | N/A | Normal |
| | H7 | C / C / C / C | 4 / 4 / 4 / 4 | 0 / 0 / 0 / 0 | ++ / ++ / ++ / ++ | 1 / 1 | N/A | Normal |
| | H8 | C / C / C / C | 4 / 4 / 4 / 4 | 0 / 0 / 0 / 0 | ++ / ++ / ++ / ++ | 1 / 1 | N/A | Normal |
| | H9 | C / C / C / C | 4 / 4 / 4 / 4 | 0 / 0 / 0 / 0 | ++ / ++ / ++ / ++ | 1 / 1 | N/A | Normal |
| | H10 | C / C / C / C | 4 / 4 / 4 / 4 | 0 / 0 / 0 / 0 | ++ / ++ / ++ / ++ | 1 / 1 | N/A | Normal |
| | H11 | C / C / C / C | 4- / 4- / 4- / 4- | 0 / 0 / 0 / 0 | ++ / ++ / ++ / ++ | 1 / 1 | N/A | Normal |
| Katona (early treatment) | K12 | C / C / I / I | 3 / 3 / 2 / 2 | 2 / 1 / 2 / 2 | +++ / +++ / +++ / +++ | 3 / 2 | 3 | Triparesis (lower extremities and upper right limb affected) |
| | K13 | C / I / C / I | 4 / 3 / 4 / 3 | 0 / 2 / 0 / 2 | ++ / +++ / ++ / +++ | 1 / 3 | 2 | Left hemiparesis |
| | K14 | C / C / C / C | 4 / 4 / 4 / 4 | 0 / 0 / 0 / 0 | ++ / ++ / +++ / +++ | 1 / 1 | 1 | Normal |
| | K15 | C / I / C / I | 4 / 3 / 4 / 3 | 0 / 2 / 0 / 2 | ++ / +++ / ++ / +++ | 1 / 2 | 2 | Left hemiparesis |
| | K16 | C / C / C / C | 4 / 3 / 4 / 3 | 0 / 0 / 0 / 0 | ++ / +++ / ++ / +++ | 1 / 2 | 1 | Monoparesis of left upper limb |
| | K17 | C / C / C / C | 3 / 4 / 3 / 4 | 1+ / 0 / 1+ / 0 | +++ / ++ / +++ / ++ | 3 / 1 | 2 | Right hemiparesis |
| | K18 | C / C / C / C | 4 / 4 / 4 / 4 | 0 / 0 / 0 / 0 | ++ / ++ / ++ / ++ | 1 / 1 | 1 | Normal |
| | K19 | C / C / C / C | 3 / 4 / 3 / 4 | 0 / 0 / 1 / 0 | +++ / ++ / ++ / ++ | 2 / 1 | 1 | Right hemiparesis |
| | K20 | C / I / C / C | 4 / 3 / 4 / 3 | 0 / 1+ / 0 / 0 | ++ / +++ / ++ / +++ | 1 / 2 | 1 | Left hemiparesis |
| | K21 | I / I / I / I | 3 / 4 / 3 / 4 | 2 / 0 / 1+ / 0 | +++ / ++ / +++ / ++ | 3 / 1 | 1 | Right hemiparesis |
| | K22 | I / C / I / I | 2 / 3 / 2 / 2 | 1+ / 0 / 2 / 1+ | ++ / ++ / +++ / +++ | Did not cooperate | 5 | Triparesis (lower extremities and upper right limb affected) |
| | K23 | I / I / I / I | 2 / 2 / 1 / 1 | 1+ / 2 / 2 / 3 | +++ / +++ / +++ / ++++ | 4 / 3 | 3 | Quadriparesis with predominance in left hemibody |
| | L24 | I / I / I / I | 3 / 2 / 1 / 1 | 1+ / 2 / 3 / 4 | ++ / +++ / +++ / +++ | 2 / 3 | 4 | Triparesis (lower extremities and upper left limb affected) |
| | L25 | I / I / C / C | 3 / 3 / 3 / 3 | 1 / 1+ / 2 / 2 | + / +++ / + / +++ | 2 / 3 | 3 | Paraparesis |
| | L26 | I / I / I / I | 2 / 3 / 1 / 2 | 3 / 1+ / 4 / 2 | +++ / +++ / +++ / +++ | 4 / 3 | 4 | Quadriparesis with predominance in right hemibody |
| **Control (late treatment)** | L27 | C / C / I / I | 3 / 3 / 3 / 3 | 0 / 0 / 0 / 1+ | ++ / +++ / ++ / +++ | 1 / 2 | 2 | Left hemiparesis |
| | L28 | C / C / I / I | 3 / 3 / 2 / 3 | 1+ / 0 / 3 / 2 | +++ / +++ / +++ / +++ | 3 / 2 | 3 | Quadriparesis with predominance in right hemibody |
| | L29 | I / I / I / I | 2 / 3 / 2 / 3 | 2 / 1+ / 2 / 1+ | +++ / +++ / +++ / +++ | 4 / 3 | 3 | Quadriparesis with predominance in right hemibody |
| | L30 | C / C / I / I | 3 / 3 / 2 / 2 | 1+ / 0 / 1+ / 1 | ++ / ++ / ++ / ++ | 3 / 2 | 4 | Quadriparesis with predominance in right hemibody |
| | L31 | C / C / I / I | 3 / 4 / 3 / 4 | 0 / 0 / 0 / 0 | +++ / ++ / +++ / ++ | 2 / 1 | 1 | Right hemiparesis |
| | L32 | I / I / I / I | 3 / 2 / 1 / 1 | 1 / 1+ / 1+ / 3 | +++ / +++ / +++ / +++ | 4 / 4 | 4 | Triparesis (lower extremities and upper left limb affected) |

GMFCS: Gross Motor Function Classification System. MACS: Manual Ability Classification System. ROM: Range of Motion.
C: Complete. I: Incomplete. N/A: Not Applicable.

**TABLE 3. Qualitative Magnetic Resonance Imaging**

| Group | | Cerebral Gray Matter Lesion | | Cerebral White Matter Lesion | | | | | | | Cerebellum Lesion | MRI Grade | Perinatal Brain Damage |
|---|---|---|---|---|---|---|---|---|---|---|---|---|---|
| | | | | Periventricular White Matter Loss | | | | PLIC Alteration | CC Alteration | Cyst(s) | | | |
| | | Cortico-subcortical Alteration | Basal Ganglia and/or Thalamus Alteration | Frontal | Parietal | Occipital | Temporal | | | | | | |
| Katona (early treatment) | K12 | Bilateral fronto-parietal cortical atrophy with left predominance and right frontal encephalomalacia in opercular region | Both thalamus diminished | Bilateral | Bilateral | Bilateral | None | None | Hypoplastic | Bilateral periventricular glial scars in both frontal and parietal regions. | Underdevelopment of both cerebellar hemispheres | Severe | Cystic PVL / right frontal infarction and left fronto-parietal atrophy |
| | K13 | Right fronto-parietal cortical atrophy | Right thalamus diminished | Right | Bilateral | Bilateral | Bilateral | Right | Hypoplastic | Right fronto-parietal porencephalic | Underdevelopment of left hemisphere | Severe | Right PVL/periventricular hemorrhagic infarction |
| | K14 | Bilateral fronto-parieto-occipital cortical atrophy with left predominance | None | None | None | None | None | None | Posterior third diminished | None | Atrophy / loss of declive, atrophy of simple lobule, atrophy of superior and inferior semilunar lobule | Severe | Bilateral (with left predominance) frontal, parietal and occipital region cortical-subcortical infarction |
| | K15 | None | Right thalamus, globus pallidus, putamen and caudate | Right | Right | None | None | Right | Middle third diminished | One cyst that spreads from the insula to the right frontal region | None | Moderate | Right PLIC infarction |

| | | | | | | | | | | | | |
|---|---|---|---|---|---|---|---|---|---|---|---|---|
| K16 | None | Right thalamus, globus pallidus, putamen and caudate | Right | None | Right (minimum) | Right | Right | Hypoplastic | Right fronto-temporal porencephalic | Underdevelopment of left hemisphere | Severe | Right periventricular hemorrhagic infarction (extensive in frontopolar and temporal regions) |
| K17 | Left fronto-parietal encephalomalacia (in inferior perirolandic region) | Right thalamus, globus pallidus, putamen and caudate | Left | Left | None | None | Left | Body and isthmus diminished | One cyst that spreads from the insula to the left fronto-parietal region | None | Moderate | Left inferior perirolandic cortex and PLIC infarction |
| K18 | Bilateral parietal region encephalomalacia with left predominance | None | None | None | None | None | None | None | None | None | Moderate | Bilateral (with left predominance) parietal infarction |
| K19 | Left fronto-parietal encephalomalacia | None | None | None | None | None | None | None | None | None | Moderate | Left perirolandic region infarction |
| K20 | Extensive lesion in right hemisphere and left parieto-occipital encephalomalacia | Severe lesion of right globus pallidus, putamen, caudate and thalamus | Right | Bilateral | Bilateral | Right | Right | Hypoplastic | Right parieto-temporal porencephalic | Underdevelopment of both cerebellar hemishperes with left predominance | Severe | Left parieto-occipital infarction and right brain hemisphere extensive infarction (occlusion of the middle cerebral and posterior cerebral right arteries) |
| K21 | Left fronto-parietal cortical atrophy | Left globus pallidum, putamen, caudate and thalamus loss | Left | Bilateral | Bilateral | Left | Left | Severely hypoplastic | Left fronto-parieto-occipito-temporal extensive porencephalic | Underdevelopment of right hemisphere | Severe | Left periventricular hemorrhagic infarction / PVL |
| K22 | Bilateral fronto-parietal | Both thalamus diminished | Bilateral | Bilateral | Bilateral | Bilateral | Bilateral with left predomina | Hypoplastic | Bilateral periventricular glial scars in | Underdevelopment of both cerebellar hemispheres | Severe | Cystic PVL |

| | | | | | | | | | | | | | |
|---|---|---|---|---|---|---|---|---|---|---|---|---|---|
| | | cortical atrophy | | | | | | nce | | parieto-occipital region | | | |
| | K23 | Bilateral fronto-parietal cortical atrophy | Both thalamus diminished | Bilateral | Bilateral | Bilateral | None | Bilateral | Hypoplastic | Bilateral periventricular glial scars in frontal and parietal regions | None | Severe | Cystic PVL |
| **Control (late treatment)** | L24 | None | Both thalamus diminished | Bilateral | Bilateral | None | None | Bilateral with right predomina nce | Hypoplastic | Bilateral periventricular glial scars in frontal and parietal regions | None | Severe | Cystic PVL |
| | L25 | Bilateral fronto-parietal cortical atrophy | Both thalamus diminished | Bilateral | Bilateral | Bilateral | None | Bilateral with left predomina nce | Hypoplastic | Bilateral periventricular glial scars in frontal, parietal and occipital regions | None | Severe | Cystic PVL |
| | L26 | Bilateral fronto-parieto-occipital cortical atrophy | Both thalamus diminished | Bilateral | Bilateral | Bilateral | None | Bilateral | Hypoplastic | Bilateral periventricular glial scars in frontal and parietal regions | Underdevelopment of both cerebellar hemispheres | Severe | Cystic PVL |
| | L27 | None | Light diminishing in both thalamus | Bilateral | Bilateral | Bilateral | None | Slightest bilateral | Diminished | Bilateral glial scars at right radiated crown level in frontal and parietal regions | None | Severe | Cystic PVL |
| | L28 | Bilateral fronto-parietal cortical atrophy | Both thalamus diminished and both lesion and diminishing in left caudate | Bilateral | Bilateral | Left | None | Bilateral with left predomina nce | Hypoplastic | Bilateral periventricular glial scars in frontal and parietal regions | Underdevelopment of both cerebellar hemispheres | Severe | Left periventricular hemorrhagic infarction / cystic PVL |
| | L29 | Bilateral fronto-parietal cortical atrophy | Severe lesion in left globus pallidum, putamen, caudate and | Bilateral with left predomi nance | Bilateral | Bilateral with left predomi nance | Bilateral | Bilateral with left predomina nce | Hypoplastic | Left frontal porencephalic and periventricular glial scars in frontal and parietal regions | Hypoplasia of both cerebellar hemispheres | Severe | Left periventricular hemorrhagic infarction / cystic PVL |

| | | | | | | | | | | | | |
|---|---|---|---|---|---|---|---|---|---|---|---|---|
| | | thalamus. Right side diminished. | | | | | | | | | | |
| L30 | Bilateral fronto-parietal cortical atrophy | Both thalamus diminished with left predominance | Bilateral | Bilateral | Bilateral | None | Bilateral with left predominance | Hypoplastic | Bilateral (with left predominance) periventricular glial scars in frontal and parietal regions | Underdevelopment of both cerebellar hemispheres | Severe | Cystic PVL |
| L31 | Bilateral frontal cortical atrophy | None | None | Left | Left | None | None | Hypoplastic | Bilateral periventricular T2 hyperintensity and T1 hypointensity in parietal and occipital regions | None | Moderate | Non-cystic PVL with left predominance |
| L32 | Bilateral fronto-parieto-occipito-temporal cortical atrophy | Both thalamus diminished | Bilateral | Bilateral | Bilateral | None | Bilateral | Hypoplastic | Bilateral periventricular glial scars in frontal, parietal and occipital regions | Underdevelopment of both cerebellar hemispheres | Severe | Cystic PVL |

CC: Corpus Callosum. MRI: Magnetic Resonance Imaging. PLIC: Posterior Limb of Internal Capsule. PVL: Periventricular Leukomalacia.

**TABLE 4. Quantitative Magnetic Resonance Imaging**

| Group | | Corpus Callosum Volume | Right Lateral Ventricle Volume | Left Lateral Ventricle Volume | Right PLIC Area | Left PLIC Area |
|---|---|---|---|---|---|---|
| **Healthy** | H1 | 10.1 $cm^3$ | 6.7 $cm^3$ | 6.5 $cm^3$ | 1.2 $cm^2$ | 1.1 $cm^2$ |
| | H2 | 10.6 $cm^3$ | 4.2 $cm^3$ | 5.4 $cm^3$ | 1.3 $cm^2$ | 1.3 $cm^2$ |
| | H3 | 14.1 $cm^3$ | 9.7 $cm^3$ | 9.5 $cm^3$ | 1.2 $cm^2$ | 1.3 $cm^2$ |
| | H4 | 14.7 $cm^3$ | 6.7 $cm^3$ | 4.9 $cm^3$ | 1.3 $cm^2$ | 1.2 $cm^2$ |
| | H5 | 11.8 $cm^3$ | 6.6 $cm^3$ | 10.8 $cm^3$ | 1.4 $cm^2$ | 1.5 $cm^2$ |
| | H6 | 14.6 $cm^3$ | 6.5 $cm^3$ | 10.9 $cm^3$ | 1.3 $cm^2$ | 1.4 $cm^2$ |
| | H7 | 10.3 $cm^3$ | 7.7 $cm^3$ | 3.7 $cm^3$ | 1.5 $cm^2$ | 1.1 $cm^2$ |
| | H8 | 12.7 $cm^3$ | 9.2 $cm^3$ | 13.2 $cm^3$ | 1.2 $cm^2$ | 1.1 $cm^2$ |
| | H9 | 12.4 $cm^3$ | 6.7 $cm^3$ | 7.5 $cm^3$ | 1.4 $cm^2$ | 1.2 $cm^2$ |
| | H10 | 9.8 $cm^3$ | 4.2 $cm^3$ | 4.9 $cm^3$ | 1.2 $cm^2$ | 1.2 $cm^2$ |
| | H11 | 13.7 $cm^3$ | 7.5 $cm^3$ | 9.5 $cm^3$ | 1.2 $cm^2$ | 1.3 $cm^2$ |
| | **Mean** | **12.3 $cm^3$** | **6.9 $cm^3$** | **7.9 $cm^3$** | **1.3 $cm^2$** | **1.2 $cm^2$** |
| | **SD** | **1.8 $cm^3$** | **1.7 $cm^3$** | **3.0 $cm^3$** | **0.1 $cm^2$** | **0.1 $cm^2$** |
| **Katona (early treatment)** | K12 | 3.9 $cm^3$ | 22.7 $cm^3$ | 16.6 $cm^3$ | 0.7 $cm^2$ | 0.7 $cm^2$ |
| | K13 | 4.9 $cm^3$ | 85.2 $cm^3$ | 40.9 $cm^3$ | 0.1 $cm^2$ | 1.0 $cm^2$ |
| | K14 | 9.7 $cm^3$ | 6.7 $cm^3$ | 6.3 $cm^3$ | 1.2 $cm^2$ | 1.1 $cm^2$ |
| | K15 | 8.1 $cm^3$ | 5.9 $cm^3$ | 5.1 $cm^3$ | 0.2 $cm^2$ | 1.0 $cm^2$ |
| | K16 | 4.0 $cm^3$ | 33.1 $cm^3$ | 7.8 $cm^3$ | 0.9 $cm^2$ | 0.9 $cm^2$ |
| | K17 | 8.7 $cm^3$ | 4.9 $cm^3$ | 6.4 $cm^3$ | 1.3 $cm^2$ | 0.1 $cm^2$ |
| | K18 | 10.6 $cm^3$ | 4.3 $cm^3$ | 4.3 $cm^3$ | 1.2 $cm^2$ | 1.0 $cm^2$ |
| | K19 | 12.2 $cm^3$ | 6.4 $cm^3$ | 7.9 $cm^3$ | 1.0 $cm^2$ | 1.0 $cm^2$ |
| | K20 | 3.3 $cm^3$ | 139.9 $cm^3$ | 9.2 $cm^3$ | 0.4 $cm^2$ | 0.9 $cm^2$ |
| | K21 | 0.2 $cm^3$ | 23.3 $cm^3$ | 240.5 $cm^3$ | N/A | N/A |
| | K22 | 9.0 $cm^3$ | 62.9 $cm^3$ | 81.1 $cm^3$ | 1.3 $cm^2$ | 1.3 $cm^2$ |
| | K23 | 3.6 $cm^3$ | 21.0 $cm^3$ | 26.0 $cm^3$ | 0.6 $cm^2$ | 0.8 $cm^2$ |
| | **Mean** | **6.5 $cm^3$** | **34.7 $cm^3$** | **37.7 $cm^3$** | **0.8 $cm^2$** | **0.9 $cm^2$** |
| | **SD** | **3.6 $cm^3$** | **41.6 $cm^3$** | **67.6 $cm^3$** | **0.4 $cm^2$** | **0.2 $cm^2$** |
| **Control (late treatment)** | L24 | 6.7 $cm^3$ | 14.9 $cm^3$ | 18.5 $cm^3$ | 0.8 $cm^2$ | 1.1 $cm^2$ |
| | L25 | 5.7 $cm^3$ | 11.9 $cm^3$ | 13.1 $cm^3$ | 1.2 $cm^2$ | 1.3 $cm^2$ |
| | L26 | 5.6 $cm^3$ | 26.7 $cm^3$ | 26.8 $cm^3$ | 0.9 $cm^2$ | 1.0 $cm^2$ |
| | L27 | 11.1 $cm^3$ | 35.5 $cm^3$ | 56.0 $cm^3$ | 1.1 $cm^2$ | 1.1 $cm^2$ |
| | L28 | 8.0 $cm^3$ | 27.6 $cm^3$ | 31.5 $cm^3$ | 0.8 $cm^2$ | 1.2 $cm^2$ |
| | L29 | 5.4 $cm^3$ | 37.1 $cm^3$ | 60.2 $cm^3$ | 0.8 $cm^2$ | 0.3 $cm^2$ |
| | L30 | 7.7 $cm^3$ | 22.3 $cm^3$ | 30.4 $cm^3$ | 1.0 $cm^2$ | 1.1 $cm^2$ |
| | L31 | 5.2 $cm^3$ | 7.5 $cm^3$ | 10.6 $cm^3$ | 0.9 $cm^2$ | 0.9 $cm^2$ |
| | L32 | 5.2 $cm^3$ | 20.1 $cm^3$ | 19.0 $cm^3$ | 0.5 $cm^2$ | 0.6 $cm^2$ |
| | **Mean** | **6.7 $cm^3$** | **22.6 $cm^3$** | **29.6 $cm^3$** | **0.9 $cm^2$** | **0.9 $cm^2$** |
| | **SD** | **1.9 $cm^3$** | **10.1 $cm^3$** | **17.7 $cm^3$** | **0.1 $cm^2$** | **0.3 $cm^2$** |

PLIC: Posterior Limb of Internal Capsule. N/A: Not Available.

**TABLE 5. Confusion Matrix for Clustering Results.**

**a) Quantitative Magnetic Resonance Imaging**

| | | Healthy | Patient | | |
|---|---|---|---|---|---|
| | | **C1** | **C2** | **C3** | |
| Healthy | **C1** | **11** | **0** | **0** | **11** |
| Patient | **C2** | **0** | **5** | **7** | **12** |
| | **C3** | **0** | **3** | **6** | **9** |
| | | **11** | **8** | **13** | |

**Output Class** (columns); **Target Class** (rows)

**b) Motor Evoked Potentials**

| | | Healthy | Patient | | |
|---|---|---|---|---|---|
| | | **C1** | **C2** | **C3** | |
| Healthy | **C1** | **10** | **1** | **0** | **11** |
| Patient | **C2** | **8** | **4** | **0** | **12** |
| | **C3** | **2** | **5** | **2** | **9** |
| | | **20** | **10** | **2** | |

**Output Class** (columns); **Target Class** (rows)

**c) Neuropediatric Evaluation**

| | | Healthy | Patient | | |
|---|---|---|---|---|---|
| | | **C1** | **C2** | **C3** | |
| Healthy | **C1** | **11** | **0** | **0** | **11** |
| Patient | **C2** | **8** | **3** | **1** | **12** |
| | **C3** | **0** | **4** | **5** | **9** |
| | | **19** | **7** | **6** | |

**Output Class** (columns); **Target Class** (rows)

Confusion matrix for clustering results based on three classes (C1 for Healthy group, C2 for early treated group with Katona therapy and C3 for late treated group without Katona therapy) of three databases: quantitative magnetic resonance imaging (a), motor evoked potentials (b) and neuropediatric evaluation / motor performance tests (c). The color green is related to healthy participants, yellow to patients (with and without Katona therapy) and red with misclassification.

**FIGURES**

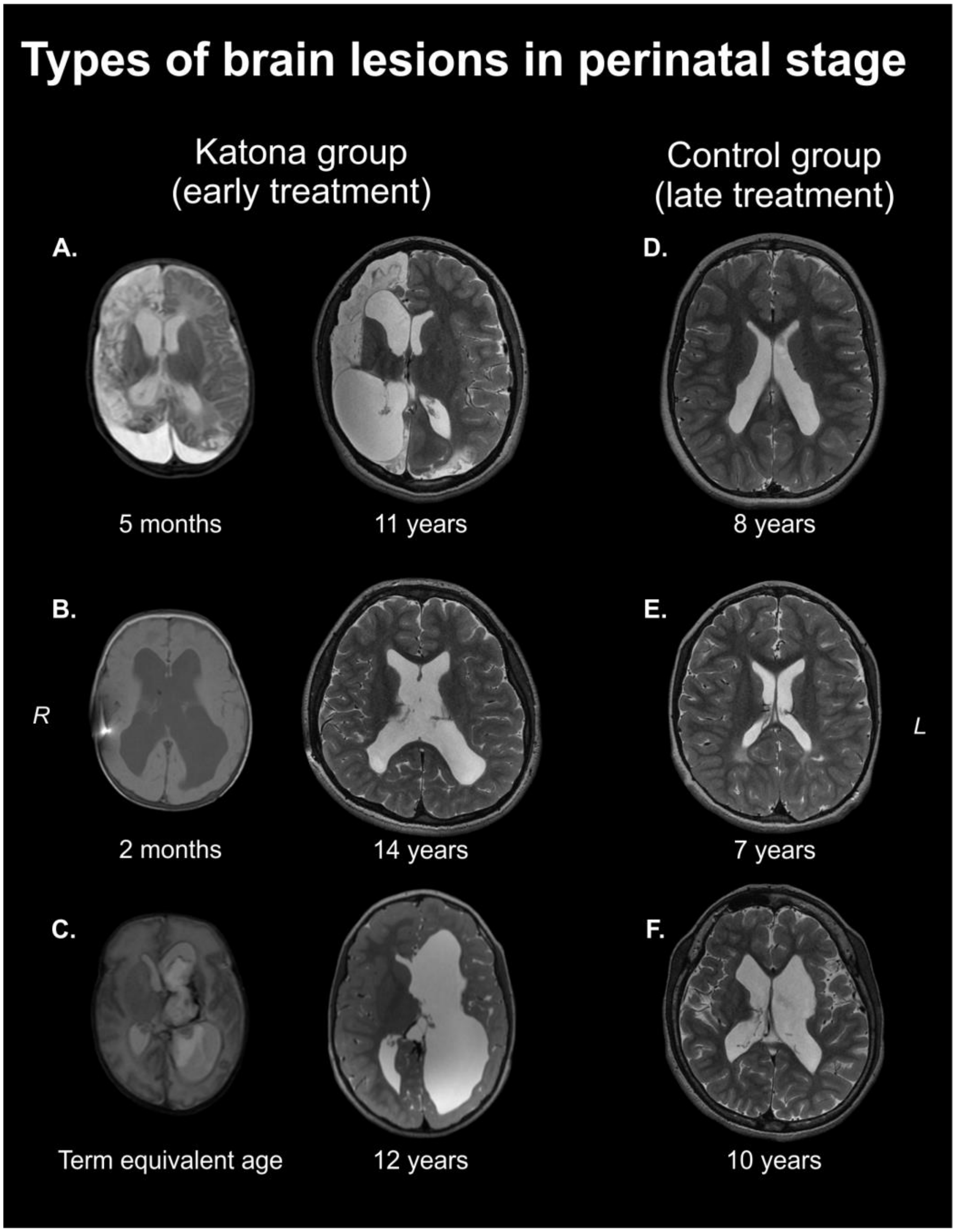

**Fig. 1.** Magnetic resonance imaging of perinatal brain damage. **A)** Arterial ischemic stroke. **B)** Cystic periventricular leukomalacia (PVL). **C)** Periventricular hemorrhagic infarction and cystic PVL. **D)** Cystic PVL. **E)** Neuroradiological findings compatible with non-cystic PVL. **F)** Cystic PVL and left frontal porencephalic cyst secondary to periventricular hemorrhagic infarction. Images in radiological convention.
*R: Right. L: Left.*

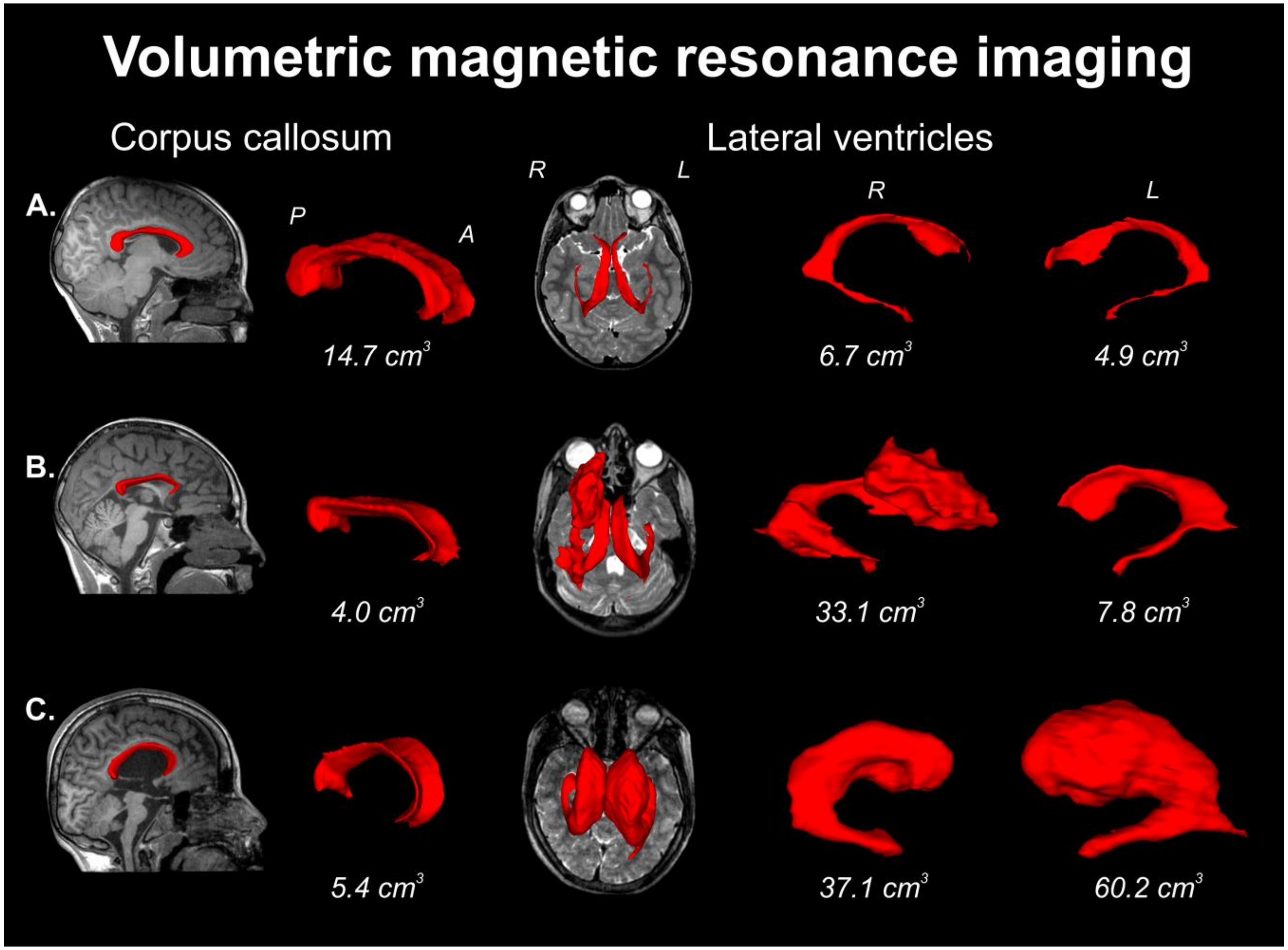

**Fig. 2.** Volumetric magnetic resonance imaging. 3D reconstruction of the corpus callosum and lateral ventricles with corresponding volumetric measurement and sagittal T1-weighted and axial T2-weighted sequences with overlaying volumetric reconstruction. **A)** Healthy child at 9 years old. **B)** Patient of the Katona group (early treated) at 11 years old. **C)** Patient of the Control group (late treated) at 10 years old.
*A: Anterior. P: Posterior. R: Right. L: Left.*

# SUPPLEMENTARY MATERIAL

## MAGNETIC RESONANCE IMAGING PARAMETERS

Brain magnetic resonance imaging (MRI) studies in infancy stage with early perinatal brain damage (PBD) detection aim were only for the Katona group (only group with early diagnosis and longitudinal follow-up) and were acquired with a 1.0 Tesla MRI scanner (Philips Medical Systems, Best, Netherlands). Infant brains were scanned during natural sleep and used ear protection. Structural images included: axial and sagittal T1-weighted conventional spin echo (SE), repetition time (TR) = 405 ms, echo time (TE) = 15 ms, flip angle 62°, slices 15, slice thickness 5 mm, matrix 256 x 166, field of view (FoV) = 220 x 220 mm, voxel sizes of 6.0 x 0.8 x 0.8 $mm^3$. Axial and coronal T2-weighted SE, TR/TE 2600/150 ms, flip angle 90°, slices 30, slice thickness 6 mm, matrix 256 x 153, FoV 200 x 200 $mm^2$, voxel sizes of 0.8 x 6.6 x 0.8 $mm^3$.

Brain MRI studies for all participants in a range 7-to-16 years (transversal analysis) were acquired with a 3.0 Tesla scanner (General Electric Healthcare, Milwaukee, Wisconsin, US). A 16-channel neurovascular head coil (HDNV) was used for data acquisition. All participants were awake and used ear protection. Structural images included: coronal 3D T1-weigthed SPGR, TR/TE 6/2 ms, flip angle 12°, slices 392, slice thickness 1 mm, matrix 224 x 224, FoV 220 x 220 $mm^2$, voxel sizes of 0.8 x 0.5 x 0.8 $mm^3$. Coronal 3D T2-weigthed SE, TR/TE 2500/68 ms, flip angle 90°, slices 196, slice thickness 1 mm, matrix 224 x 224, FoV 220 x 220 $mm^2$, voxel sizes of 0.8 x 1.0 x 0.8 $mm^3$. Axial T2-weigthed FSE, TR/TE 7000/100 ms, flip angle 111°, slices 70, slice thickness 2 mm, matrix 448 x 352, FoV 220 x 220 $mm^2$, voxel sizes of 0.4 x 0.4 x 2.0 $mm^3$. Axial 3D-TOF SPGR, TR/TE 20/3 ms, flip angle 15°, slices 120, slice thickness 1.2 mm, matrix 352 x 192, FoV 180 x 180 $mm^2$, voxel sizes of 0.3 x 0.3 x 0.6 $mm^3$. Diffusion tensor spin echo, single-shot EPI, TR/TE 8000/100 ms, flip angle 90°, slices 50, slice thickness 2 mm, matrix 128 x 128, FoV 220 x 220 $mm^2$, voxel sizes of 0.8 x 0.8 x 2.0 $mm^3$. We obtained diffusion-weighted images along 64 different directions with a b-value of 2000 $s/mm^2$, and 2 non-diffusion-weighted images ($b = 0$ $s/mm^2$). Time acquisition for these 5 sequences was approximately 22 min.

# METHODOLOGY FOR THE QUANTITATIVE MAGNETIC RESONANCE IMAGING ANALYSIS

## Volumetric analysis

Volumetric analysis of the corpus callosum and lateral ventricles was performed from coronal 3D T1-weighted SPGR sequence. The T1-weighted sequence allows a better observation between the white matter and gray matter contrast, favoring the manual segmentation of the corpus callosum. The manual segmentation, quantitative analysis and 3D models visualization were performed with 3D slicer v4.10.2 software.

### 1. Manual segmentation

#### 1.1 Initial inspection and image orientation for segmentation

The anatomical boundaries and morphological details are not easily determined in a single plane. Sometimes the visualization only in one plane (*e.g.,* sagittal for corpus callosum) is not enough to determine the anatomical limits. It is necessary to visualize the three planes (sagittal, axial and coronal) to endorse an adequate marking and an optimal visualization of the lateral, superior, inferior, anterior and posterior limits. Therefore, visual evaluation of the sagittal, coronal and axial plane is made to determine the anatomical boundaries of the corpus callosum and lateral ventricles and/or to be able to detect any anatomical abnormality that may interfere with the final volumetry. Afterwards, the operator positions itself systematically in a specific plane to initiate the segmentation process. The initial orientation of the corpus callosum is determined in the sagittal plane (at the anterior commissure – posterior commissure line level), whereas the initial orientation of the lateral ventricles is determined by the axial plane at the ventricular dome height (at superior corona radiata level).

### 1.2 Manual segmentation direction

The corpus callosum segmentation is made in the sagittal plane in a bilateral parasagittal direction using as a secondary anatomical reference the axial plane. For the lateral ventricles' segmentation, the axial plane is used with a cranial-caudal direction and with the sagittal and coronal planes as secondary anatomical reference.

## 2. Anatomical boundaries

### 2.1 Corpus callosum

The anatomical boundaries of the corpus callosum are not clear in the parasagittal plane. In general terms, conventional magnetic resonance imaging (MRI) don't allow differentiation of the fascicles. Consequently, the lateral portions of the corpus callosum are determined in the axial plane to corroborate that the corpus callosum is not being overestimated when the cingulum is included in planes where the corpus callosum limits are undistinguishable (parasagittal). The information that the axial plane gives is fundamental for delimiting the lateral anatomical boundaries of the corpus callosum. The anatomical point of reference in the axial plane at the genu level is the caudate head and the cortex of the cingulum gyrus (at frontal lobe level). The anatomical point of reference at the splenium level in the axial plane is the gyrus of the cingulum (at parietal level) and the lateral portion of the choroid plexus of the lateral ventricles. The superior, inferior, anterior and posterior anatomical boundaries of the corpus callosum are clear in the sagittal plane thanks to the contrast between white and gray matter of the T1- weighted. These anatomical boundaries are given by the marginal callous fissure and cingulum gyrus. Systematically, the corpus callosum is segmented including the genu, body, isthmus and splenium (not including the tapetum).

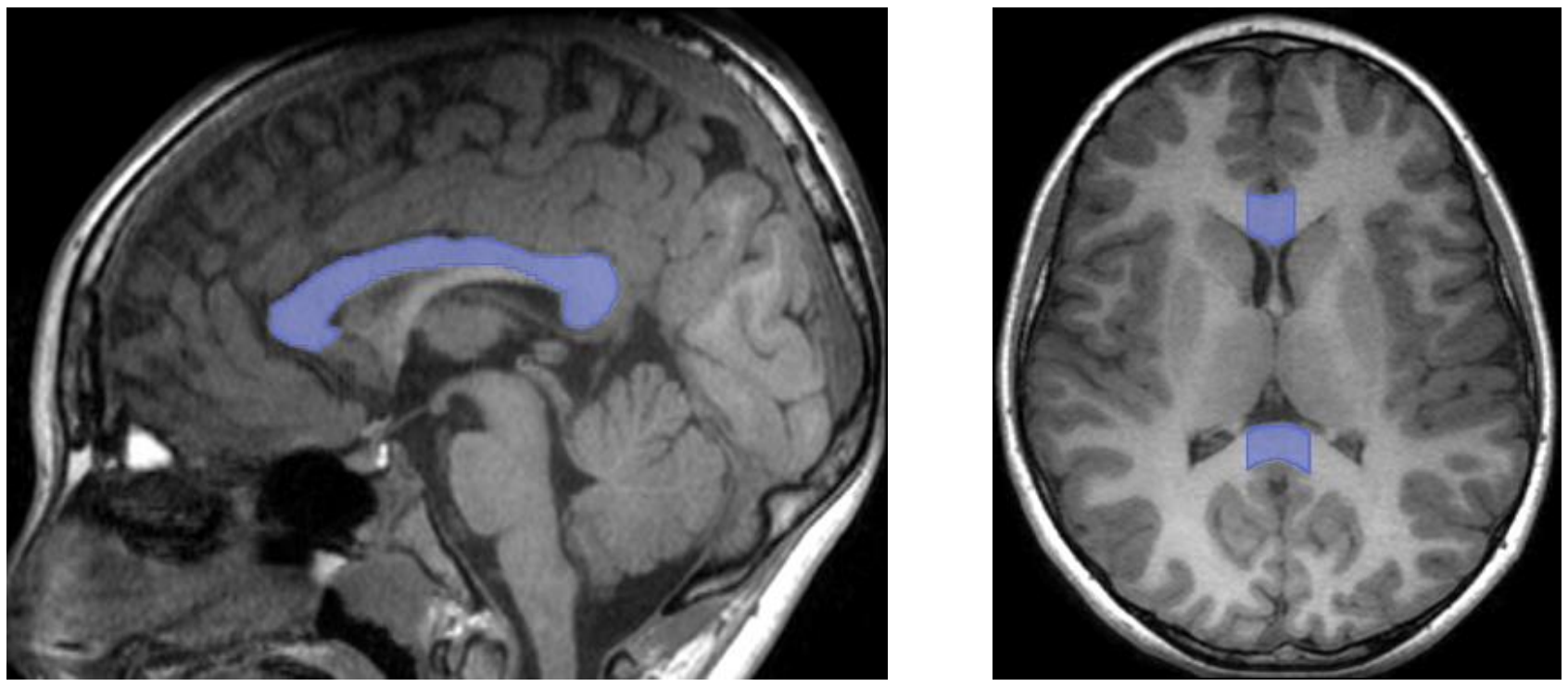

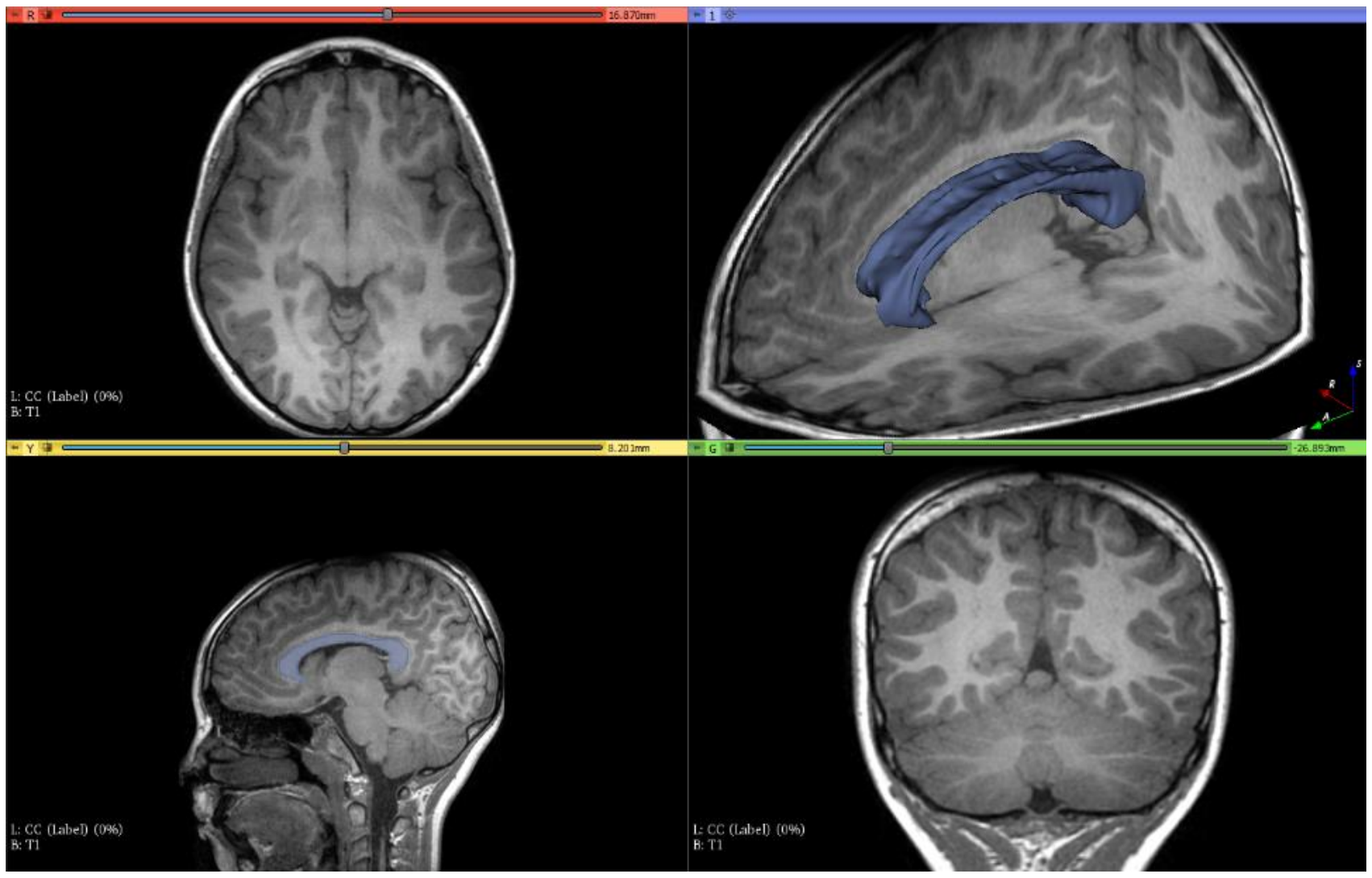

### 2.2 Lateral ventricles

The anatomical boundaries of the lateral ventricles are clear despite its morphological variability in pediatric stage. The manual segmentation of the lateral ventricles includes the frontal horn, the body of lateral ventricle and occipital and temporal horn. The lateral anatomical boundaries in dorsal-ventral direction are the caudate nucleus and thalamus at the ventricular dome / body level, and corresponding periventricular white matter. The medial boundaries in cranial-caudal direction are the septum pellucidum, columns (anterior pillars) and body of the fornix, interventricular foramina of Monro and the crura of the fornix (posterior pillars), and the hippocampus at temporal level. Any structural condition or lesion with loss of periventricular white matter that communicates widely with the lateral ventricle was measured and quantified as part of the lateral ventricle (*e.g.*, porencephalic cyst as a consequence of periventricular hemorrhagic infarction).

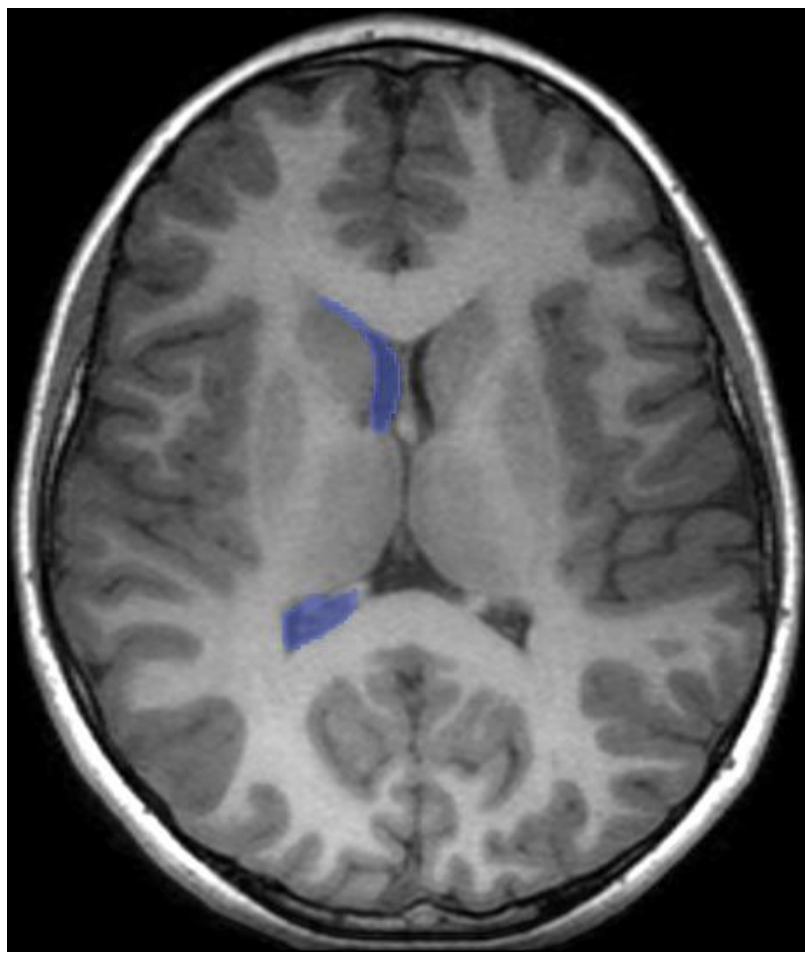

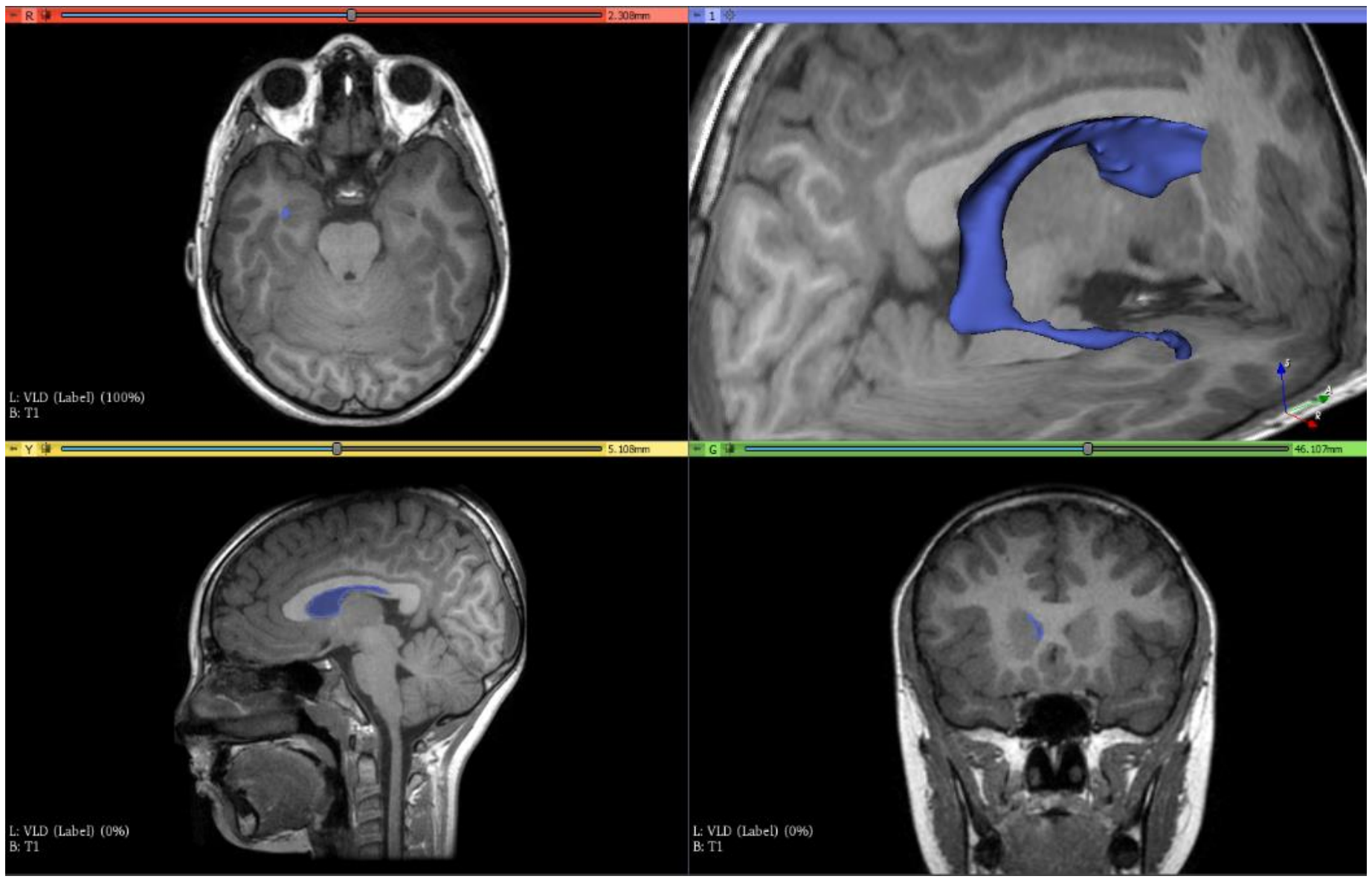

**Area of the posterior limb of internal capsule**

The areas of the posterior limb of internal capsule were calculated in RGB-color maps (color by orientation mode) from diffusion weighted imaging dataset. The RGB-color maps allows us to distinguish the major fibers / pathways of white matter and it is extremely helpful to determine the portions of the internal capsule (*e.g.,* anterior limb can be observed in green color; posterior limb can be observed in blue color). The hand-drawn of regions of interest (ROIs) were performed systematically in posterior limb of internal capsule (in blue color) in axial plane at the level of the basal ganglia. The image processing, ROIs drawings and areas calculations were performed in 3D slicer v4.10.2 software.

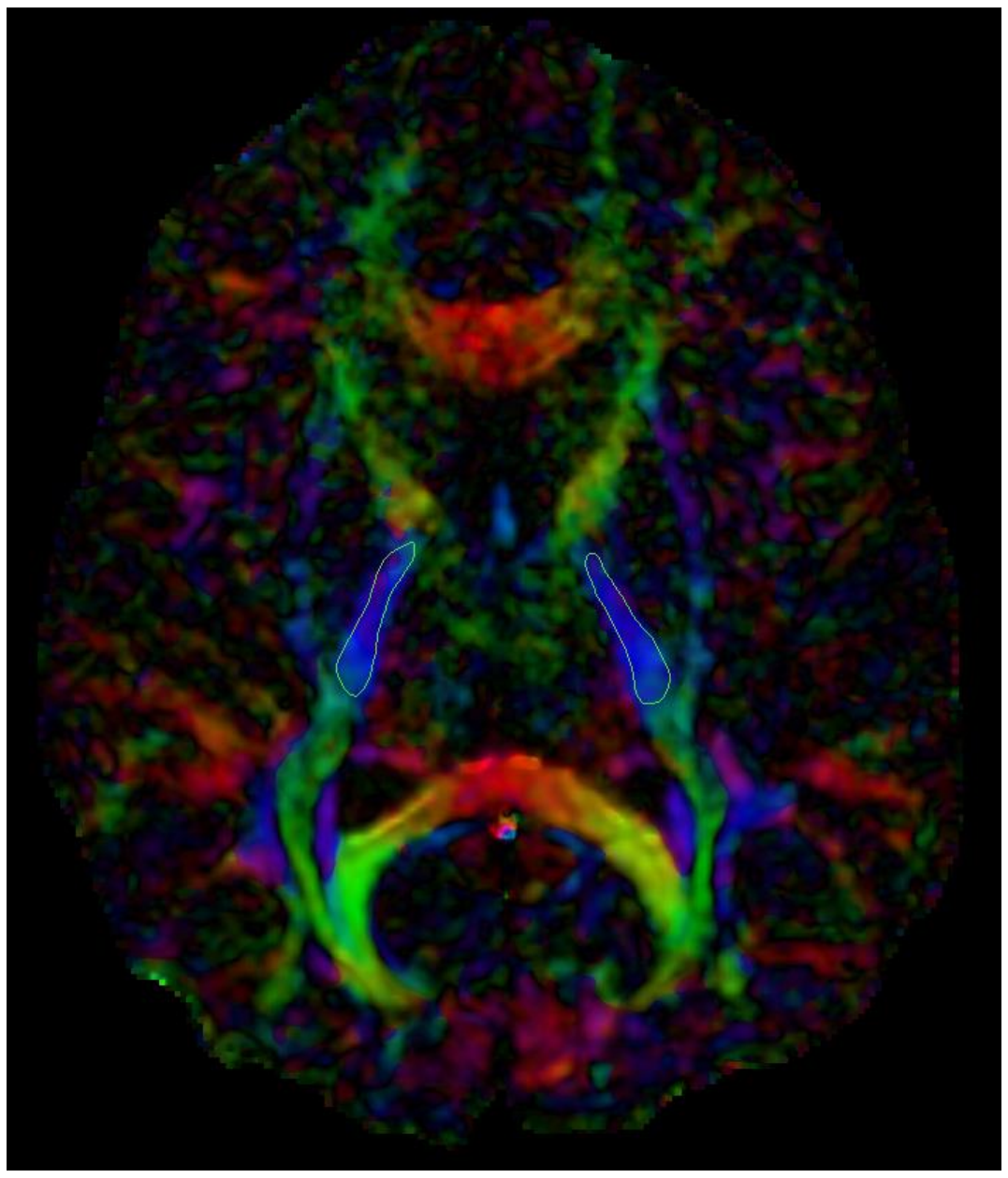